\documentclass[prd,aps,12pt]{revtex4}
\usepackage{epsf}
\def\be{\begin{equation}}
\def\ee{\end{equation}}
\def\ba{\begin{eqnarray*}}
\def\ea{\end{eqnarray*}}
\def\half{\frac{1}{2}}
\def\A{{\cal A}}
\begin{document}

\title{Massive Schwinger model with a finite inductance: 
theta-(in)dependence, the $U(1)$ problem, and low-energy theorems}
\author{S. Khlebnikov}
\affiliation{Department of Physics, Purdue University, West Lafayette, IN 47907, USA}
\begin{abstract}
Gauge theories embedded into higher-dimensional spaces with certain topologies acquire
inductance terms, which reflect the energy cost of topological charges accumulated in
the extra dimensions.
We compute topological susceptibility in the strongly-coupled two-flavor massive
Schwinger model with such an inductance term and find that it vanishes,
due to the contribution of a global low-energy mode (a ``global axion''). 
This is in accord with
the general argument on the absence of $\theta$-dependence in such topologies. 
Because the mode is a single oscillator, there is no corresponding particle,
and the solution to the $U(1)$ problem is unaffected.
\end{abstract}

\maketitle

\section{Introduction}
\label{sect:introduction}
An important aspect of the nontrivial vacuum structure of QCD 
\cite{Belavin:1975fg,'tHooft:1976up,'tHooft:1976fv,Callan:1976je,Jackiw:1976pf} is
the formation of coherent superpositions known as $\theta$-vacua.
To each value $-\pi < \theta \leq \pi$,
there corresponds a separate sector of the Hilbert space, and transitions between 
different sectors are prohibited by a selection rule. For all values of $\theta$
except $0$ and $\pi$, the theory breaks CP symmetry; the absence 
(or unnatural smallness) of this CP breaking in practice presents a naturalness 
problem, known as the strong CP problem.

The presence of a $\theta$-angle reflects the perfect degeneracy between 
states connected by topologically nontrivial (``large'') gauge transformations
\cite{Callan:1976je,Jackiw:1976pf}.
The axion solution \cite{Peccei:1977hh,Peccei:1977ur} does not destroy
this degeneracy but screens the value of  $\theta$  in a way somewhat 
similar to how the usual Higgs mechanism screens an electric 
field \cite{Dvali:2005an}. This solution gives rise to a new light particle---the 
axion \cite{Weinberg:1977ma,Wilczek:1977pj} 
(for a recent review, see Ref. \cite{Peccei:2006as}).

On the other hand, if there are extra dimensions with
suitable topology, the degeneracy is no longer protected by gauge invariance,
and that gives a reason to think that in this way
an axion-free solution to the strong CP problem can be achieved
\cite{Khlebnikov:1987zg,Khlebnikov:2004am}. 
(Other extra-dimensional solutions have been proposed in Refs. 
\cite{Aldazabal:2002py,Bars:2006dy}.)

One may worry, however, that the solution proposed in 
Refs. \cite{Khlebnikov:1987zg,Khlebnikov:2004am} is 
in conflict with the low-energy theorems of QCD, such
as those derived in Refs.
\cite{Crewther:1977ce,Shifman:1979if,Crewther:1979pi}. These theorems connect
the existence of observable $\theta$-dependencies to a successful solution to the
$U(1)$ problem---the absence of a light $\eta'$ meson. In the present paper,
we argue that this concern is not justified, using the massive 
two-flavor Schwinger model (two-dimensional electrodynamics), about which much is
known \cite{Coleman:1976uz}, as an example.

The strongly-coupled Schwinger model provides a rather close analogy to QCD, 
including a solution
to the $U(1)$ problem. Of course, not all observable quantities of QCD have direct 
analogs in this model. An important one that does is the vacuum topological 
susceptibility, a low-energy relation for which was derived in 
Refs. \cite{Crewther:1977ce,Shifman:1979if}. Here we consider the
essentially identical relation applicable in the Schwinger model.

The single (but crucial) new ingredient that we add to the model is a {\em finite
inductance}, which reflects the energy of magnetic flux in the extra dimensions.
Such a flux is the simplest example of a topological charge (instanton number),
and a finite energy associated with accumulation of the topological charge in the
extra dimensions appears naturally in the scenario 
of Ref. \cite{Khlebnikov:2004am}.

\begin{figure}
\leavevmode\epsfysize=2in \epsfbox{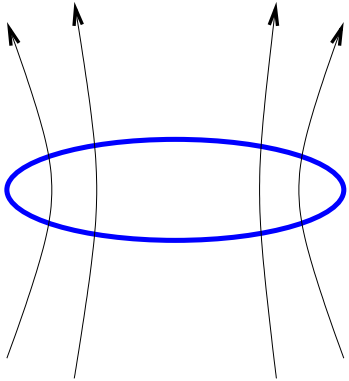}
\vspace*{0.2in}
\caption{A ring threaded by a magnetic flux. Instanton transitions
in the gauge theory on the ring correspond to changing the value of the flux,
and $\theta$-vacua---to a steady current of flux into the ring. One expects that
the energy cost associated with increasing the flux (a finite inductance) will 
prevent such steady currents.}
\label{fig:loop}
\end{figure}

Thus, we consider the theory on a ring (Fig. \ref{fig:loop}) 
embedded in a higher-dimensional space. The length $L$ of the ring is taken to
be much larger than the inverse of either the fermion mass $m$ 
or the gauge coupling $g$. The fermions 
are confined to move along the ring, but the gauge field can penetrate some distance
into the bulk.
Nevertheless, the gauge field is assumed localized, in the sense that the total number
of modes available to it
at low energies is the same as in the two-dimensional 
(one space and one time) theory. (In particular, the Coulomb law is the same as 
in two dimensions.) The precise mechanism of localization will not be important
here. (For a variant of such a mechanism, see Ref. \cite{Khlebnikov:2004am}.)

A finite inductance lifts the degeneracy between vacua corresponding to different
values of magnetic flux through the ring, and it stands to reason that 
this will prevent formation of $\theta$-vacua. We stress that the system is not
confined to a vicinity of a specific value of the flux: transitions between different
values are fully allowed, and the true vacuum is a superposition of 
different flux states. It is just that this superposition is now unique and is no 
longer characterized by a value of ``quasimomentum'' (a $\theta$-angle).

In the present paper we show that this $\theta$-independence is 
in complete accord with the low-energy theorem and
does not in any way upset the solution to the $U(1)$ problem.
What happens is that the global topological mode, which was originally frozen by the
superselection rule, is now liberated and contributes precisely the right amount to
satisfy the theorem. 

These results are derived in Sects. 2 and 3. Considerations of Sect. 2 are general,
while in Sect. 3 we restrict our attention to the strong-coupling limit,
where the analogy with QCD applies. In that limit, we explicitly compute 
the topological susceptibility and find that it is zero.

The topological mode is a single oscillator: there is no particle associated with 
it (it is a ``global axion''
in the terminology of Ref. \cite{Khlebnikov:2004am}). Thus, the price one pays for
such a solution to the strong CP problem is a (weak) violation of Lorentz invariance.

\section{The role of finite inductance}
The model contains two species of fermions of equal masses $m$, interacting with
an Abelian gauge field $A_\mu$, $\mu=0,1$. The action is
\be
S = \int dx dt \left\{ \half F_{01}^2 - \frac{g\theta}{2\pi} F_{01}
+ \sum_{f=1,2} \bar{\psi}_f [i \gamma^\mu (\partial_\mu - i g A_\mu) - m] \psi_f
\right\} - \frac{1}{2I} \int dt \Phi^2 \; ,
\label{S}
\ee
where $F_{01} = \partial_0 A_1 - \partial_1 A_0$, $\Phi$ is the magnetic flux through
the ring:
\[
\Phi = \int_0^L dx A_1 \; ,
\]
and $I$ is a finite inductance. The usual two-flavor massive Schwinger model 
is recovered in the limit $I \to \infty$.

Following Ref. \cite{Coleman:1976uz}, we will make use of the bosonization technique.
However, we will need to bosonize on a circle, as opposed to the infinite line.
Bosonization on a circle has a sizable literature on the particle-theory side of
the problem (see Ref. \cite{Hosotani:1998za} and references therein) 
and an even larger one on the condensed-matter side 
(see the review \cite{Delft&Schoeller}). For developments along different
lines, see Ref. \cite{Smilga:1996pi} (on the relation 
to conformal field theory) and Ref. \cite{Fukaya:2004kp} (on lattice results).

We summarize key bosonization formulas in Appendix A, where we also indicate 
simplifications obtained by restricting to the sector with zero total fermionic charges 
(the only sector relevant to the present problem). In particular, the crucial relation 
for the fermionic currents, in terms of two scalars $\phi_{1,2}$, in the zero-charge 
sector does not acquire any $1/L$ corrections and reads exactly as on the infinite line:
\be
\bar{\psi}_f \gamma^\mu \psi_f = 
\frac{1}{\sqrt{\pi}} \epsilon^{\mu\nu} \partial_\nu \phi_f \; .
\label{cur}
\ee
(A gauge-invariant regularization of the left-hand side is discussed in the Appendix.)

We now fix the gauge $\partial_1 A_1 = 0$, substitute Eq. (\ref{cur}) into 
Eq. (\ref{S}), and separate the action into two parts: the part containing 
the dependence on the spatially constant (``zero'') mode of $A_1$ and 
the remainder. The first of these equals
\be
S_A = L \int dt \left\{ \half (\partial_0 A_1)^2 
- \frac{g\theta}{2\pi}  \partial_0 A_1
- \mu\partial_0 \varphi  A_1 - \frac{L}{2I} A_1^2  \right\} \; ,
\label{SA}
\ee
where $\mu = g \sqrt{2/\pi}$, and $\varphi$ denotes the zero mode of 
$\phi_+=(\phi_1+\phi_2)/\sqrt{2}$ ($L$ is the length of the ring). 

Upon integrating out $A_1$, Eq. (\ref{SA}) gives rise to the following term in the
effective action for $\varphi$:
\be
S_1 = \half L \mu^2 \int dt dt' \varphi'(t) G(t-t') \varphi'(t')
\label{S1}
\ee
where $\varphi' = \varphi - \theta/2\sqrt{2\pi}$, and
\[
G(t) = \int \frac{d\omega}{2\pi}  e^{-i\omega t}
\frac{\omega^2}{\omega_{LC}^2 - \omega^2} \; ;
\]
$\omega_{LC} = \sqrt{L/I}$ is the $LC$ frequency. (The terminology comes about
because the first term in Eq. (\ref{SA}) can be thought of as
capacitive energy.)

We can now appreciate the role of finite inductance. If we formally set $I$ to infinity,
Eq. (\ref{S1}) becomes a mass term for $\varphi'$. This mass term pulls $\varphi$ towards
the value $\theta/2\sqrt{2\pi}$, and that value then shows up in the remainder of the
action. This is the source of the $\theta$-dependence described in 
Ref. \cite{Coleman:1976uz}. On the other hand, if keep $\omega_{LC}^2$ is finite and 
consider the modes with $\omega^2 \ll \omega_{LC}^2$, 
Eq. (\ref{S1}) is not a mass but a kinetic term. As a result, $\theta$ disappears from 
the theory. Its role is taken over by the initial value of $\varphi$.

As we will see in the next section, interaction with $\phi_-=(\phi_1 - \phi_2)/\sqrt{2}$ 
produces an effective
potential for $\varphi$ with a minimum at $\varphi = 0$. If $\varphi$ has a chance to
relax to this minimum (either through a decay into light particles or through friction
effects at a finite temperature), the memory of its initial value will be lost,
and no counterpart of $\theta$ will be left.

This conversion of the $\theta$-angle into a global time-dependent degree of 
freedom (a ``global axion'') is
the essence of the solution to the strong CP problem proposed
in Ref. \cite{Khlebnikov:2004am}. The solution does not seem to do much damage
to the known QCD phenomenology, and so it is natural to think that the low-energy
theorems will somehow work out correctly. To help alleviate possible doubts,
we explicitly address here the fate of one of those theorems---the relation for 
the topological susceptibility \cite{Crewther:1977ce,Shifman:1979if}.

\section{Topological susceptibility}
We now restrict ourselves to the strong coupling limit $g \gg m$, the one that provides
an adequate analogy to QCD. In this
case, the non-zero modes of $\phi_+$ are heavy, with a mass of order $g$, 
and decouple at low energy. The low-energy theory contains the light field 
$\phi_-$, described by the Hamiltonian \cite{Coleman:1976uz}
\be
H = N_m \int dx \left\{ \half p_-^2 + \half (\partial_1 \phi_-)^2 
-2 cm^{3/2} \mu^{1/2} \cos(\sqrt{2\pi}\varphi) \cos(\sqrt{2\pi}\phi_-) \right\} \; ,
\label{H}
\ee
and the zero mode $\varphi$ of $\phi_+$, described by the effective action 
[cf. Eq. (\ref{S1})]
\be
S_0 + S_1 = \half L \int \frac{d\omega}{2\pi} \left\{1  + 
\frac{\mu^2}
{\omega_{LC}^2 -\omega^2} \right\} \omega^2 |\tilde{\varphi}(\omega)|^2 \; .
\label{sum}
\ee
In this section, we compute topological susceptibility of the vacuum with $\theta=0$;
this will determine the response of that vacuum to a small $\theta$-term. Accordingly,
$\tilde{\varphi}$ in Eq. (\ref{sum}) is simply the Fourier transform of $\varphi$.
In Eq. (\ref{H}), $N_m$ denotes normal ordering with respect to the mass $m$, and $c$
is a numerical constant inherited from the bosonization procedure.

We will see that fluctuations of $\varphi$ in the ground and low excited states 
are small, so in these states $\cos(\sqrt{2\pi}\varphi)$ deviates little from unity. 
For our purposes, it will be sufficient to consider only the terms of the
zeroth and second order in $\varphi$. The first of these is the low-energy limit of
the usual two-flavor massive Schwinger model and determines the properties of the
$\phi_-$ subsystem, in particular, the ``quark condensate''
\be
C = - \langle \sum_f \bar{\psi}_f \psi_f \rangle = 
\langle 2c m^{1/2} \mu^{1/2} N_m \cos (\sqrt{2\pi} \phi_-) \rangle \; .
\label{cond}
\ee
Dependence of $C$ on $m$ can be understood by normal reordering 
with respect to the mass $M = [2c m \mu^{1/2} ]^{2/3}$,
which then becomes the only mass parameter in the Hamiltonian of $\phi_-$
\cite{Coleman:1976uz}. As a result,
\[
m C =  \langle  M^2 N_M \cos (\sqrt{2\pi} \phi_-) \rangle \sim M^2 \; ,
\]
which scales as $m^{4/3}$.

The second-order term
\[
H_2 = 
2\pi cm^{3/2} \mu^{1/2} \varphi^2 \int dx N_m \cos(\sqrt{2\pi}\phi_-) 
\]
is an interaction between $\varphi$ and $\phi_-$. It affects little $\phi_-$ but 
significantly $\varphi$: averaging it over the vacuum of $\phi_-$ gives rise
to an effective potential
${\cal E}(\varphi) = \pi m C \varphi^2$. The full effective action 
of $\varphi$ in this approximation is quadratic:
\be
S_{\rm eff} = S_0 + S_1 - L \int dt {\cal E}(\varphi) \; ,
\label{Seff}
\ee
where $S_0+ S_1$ is given by Eq. (\ref{sum}). 

We can now read off the frequency of small oscillations of $\varphi$:
\[
\omega_0^2 \approx {\cal E}''(0) \left( 1 + \frac{\mu^2}{\omega_{LC}^2} \right)^{-1} \; .
\]
The approximation sign corresponds to the limit $\omega_0 \ll \omega_{LC}$, 
which as we see is always justified. 
Depending on the relation between $\omega_{LC}$ and $\mu$, $\omega_0$ can be either much
smaller than $M$ or of the same order. It is important to note that it always belongs
to the low-energy sector.

Eq. (\ref{Seff}) also shows that the strength 
of zero-point fluctuations of $\varphi$ is controlled by the parameter
\be
\alpha = \frac{\omega_0}{L {\cal E}''(0)}  \; .
\label{alpha}
\ee
Given that $L$ is macroscopic, this parameter is always small, $\alpha \ll 1$.
This justifies the expansion in powers of $\varphi$. 
We will find, however, that, although the fluctuations of $\varphi$ are suppressed 
by a power of $L$, their contribution to the topological susceptibility is not.

Topological susceptibility is defined as
\[
K = i \int  
\langle T \frac{2g}{\pi} F_{01}(x) \frac{2g}{\pi} F_{01}(0) \rangle d^2 x\; ,
\]
where $T$ denotes the $T$-product, and the averaging is
over the true vacuum of the entire system. $K$ determines the average electric field 
in the vacuum induced by a small $\theta$-term in the Lagrangian.
It satisfies a relation virtually identical to the relation derived in Refs. 
\cite{Crewther:1977ce,Shifman:1979if} for the topological susceptibility in QCD.
In our case, the relation is
\be
K = 4m \langle \sum_f \bar{\psi}_f \psi_f \rangle + 
4i m^2 \int \langle T O(x) O(0) \rangle d^2 x \; ,
\label{K}
\ee
where $O = i  \sum_f  \bar{\psi}_f \gamma_5 \psi_f$ is the pseudoscalar density, 
$\gamma_5=\gamma^0 \gamma^1$. 
Derivation of Eq. (\ref{K})
is completely parallel to that in the case of QCD and will not be repeated here.

The correlator appearing in Eq. (\ref{K}) can be expressed as
\be
i \int \langle T O(x) O(0) \rangle d^2 x = 
2 L \sum_s \frac{1}{E_s} |\langle s | O(0) | 0\rangle |^2 \; ,
\label{sec}
\ee
where $|0\rangle$ is the vacuum of the system, and $|s\rangle$ are excited 
states, with energies $E_s$ relative to the vacuum.
The operator $O$ has the quantum numbers of an $\eta'$ meson. One relevant quantum 
number is G-parity \cite{Coleman:1976uz}, 
under which $\phi_-$ is even, while $\phi_+$ and $O$ are odd.
If there were no low-energy excitations with odd G-parity, the second term in 
Eq. (\ref{K})
would scale as $m^2$ at small $m$. Meanwhile, the first term scales as $m^{4/3}$
(the corresponding term in QCD scales linearly with the quark masses). 
Hence, the first term would win and make $K$ nonzero.

We have seen, however, that in the presence of a finite inductance there is a 
low-energy oscillator mode, $\varphi$. It has precisely the right quantum numbers
and therefore needs to be included in (\ref{sec}).
Because this mode is a single oscillator, it does not correspond to a new particle, 
and so the solution to the $U(1)$ problem (the absence of a light $\eta'$) 
is unaffected.

The pseudoscalar density can be bosonized in the usual way:
\[
i \sum_f  \bar{\psi}_f \gamma_5 \psi_f = 
2 cm N_m \cos(\sqrt{2\pi} \phi_-) N_m \sin(\sqrt{2\pi} \phi_+) \; ,
\]
but the sine needs to be reordered, to get rid of the tadpoles of the heavy
field $\phi_+$ (cf. Ref. \cite{Coleman:1976uz}). Upon doing that,
\[
i \sum_f  \bar{\psi}_f \gamma_5 \psi_f \approx 2 c m^{3/2} \mu^{1/2} 
\sin(\sqrt{2\pi} \varphi) N_m \cos(\sqrt{2\pi} \phi_-) 
\approx -\sqrt{2\pi} \varphi  \sum_f  \bar{\psi}_f  \psi_f \; ,
\]
where we have used the fact that fluctuations of $\varphi$ are small. 
Thus, the sum in (\ref{sec}) is
saturated by the first excited state of $\varphi$:
\[
i \int \langle T O(x) O(0) \rangle d^2 x = \frac{2\pi L \alpha}{\omega_0} C^2 \; ,
\]
and using Eq. (\ref{alpha}) for $\alpha$ we see that the second term in
Eq. (\ref{K}) precisely cancels the first. We obtain $K=0$, our main result.

\section{Conclusion}
The overall picture that emerges from the present study is of the world as a
giant resonator, whose global oscillating mode is coupled to a continuum 
branch (the familiar elementary particles). In the massive two-flavor Schwinger model,
analytical methods take us quite far, even in the strongly coupled limit, which
has many analogies to QCD. For that case, we have
explicitly computed the topological susceptibility
and seen that the global oscillator contributes just so as to banish
the vacuum $\theta$-dependence. 

We find it remarkable that the cancellation of the topological susceptibility 
occurs even though the coupling of the global mode to the continuum is
suppressed by the ``size of the universe'' (the length of the ring in our
example).
This mode has the quantum numbers of $\eta'$ but, being a single oscillator,
does not correspond to a new particle. Therefore, the solution to the $U(1)$ problem 
is not affected.

The author thanks M. Shaposhnikov and A. Vainshtein for discussions.
This work was
supported in part by the U.S. Department of Energy through Grant
DE-FG02-91ER40681 (Task B). 

\appendix
\section{Bosonization on a circle}
We follow the approach popular in the condensed-matter literature
(for a review, see Ref. \cite{Delft&Schoeller}, whose notation we mostly follow).
In this approach,
Bose operators are constructed
in terms of the operators of a massless free Fermi field. Since these are 
equal-time operator relations, they are true regardless of what the actual fermion
dynamics is (i.e., even if the fermion is massive and interacting, as in our case). 
In the
particle-theory language, this approach corresponds to normal ordering (a
gauge-invariant version thereof) with respect to 
a mass of order $1/L$. Once all the requisite operators are constructed in this way,
they can be reordered with respect to any mass by using the reordering relations of
Ref. \cite{Coleman:1974bu}. To simplify notation, we set $L=2\pi$.

We use the chiral representation of the $\gamma$-matrices, in which they are
$\gamma^0 = \sigma_1$ and
$\gamma^1 = - i \sigma_2$, so that $\gamma_5= \gamma^0 \gamma^1 = \sigma_3$.
The upper and lower components of a Dirac fermion are expanded in Fourier series
as follows
\[
\psi = \left( \begin{array}{l} \psi_R \\ \psi_L \end{array} \right) =
\frac{1}{\sqrt{2\pi}} \left( \begin{array}{l} \sum_k \alpha_k e^{ikx} \\
\sum_k \beta_k^\dagger e^{ikx} \end{array} \right) \; .
\]
The operators $\alpha$ and $\beta$ obey the usual equal-time anticommutation
relations. 
If $\psi$ really were a free massless fermion, these operators would correspond
to right- and left-movers. The momentum $k$ takes the values
\[
k = n_k - \delta \; ,
\]
where $n_k$ are integers, and $0\leq \delta < 1$ is a fraction depending on the
boundary conditions. We will consider a general $\delta$, even though 
compactifications that lead to theories with finite inductances typically produce
a special case---antiperiodic fermions ($\delta = \half$) \cite{Khlebnikov:2004am}.

In a gauge theory, we identify $\psi$ with the fermion in the Coulomb gauge
$\partial_1 \A_1 = 0$. So, in what follows, only the ``large'' gauge transformations,
which change $\A_1$ by a constant, will be of relevance. (In this Appendix, we use 
the calligraphic $\A$ to denote the gauge field.)

Define Bose operators ($q$ is a {\em positive} integer)
\ba
A_q & = & - \frac{i}{\sqrt{q}}  \sum_k \alpha^\dagger_{k-q} \alpha_k \; , \\
B_q & = & - \frac{i}{\sqrt{q}}  \sum_k \beta^\dagger_{k-q} \beta_k \; .
\ea
$A_q$ and $A_{q'}^\dagger$ (and $B_q$ and $B_{q'}^\dagger$) obey the usual 
Bose commutation relation. Then, the density operators are
\ba
\sum_{q,k} e^{iqx} :\alpha^\dagger_{k-q} \alpha_k: & = & 
i \sum_{q>0} \sqrt{q}
\left\{ A_q e^{iqx} - A_q^\dagger e^{-iqx} \right\} + N_\alpha \; , \\
\sum_{q,k} e^{iqx} :\beta_{k-q} \beta_k^\dagger: & = & 
i \sum_{q>0} \sqrt{q}
\left\{ B^\dagger_q e^{iqx} - B_q e^{-iqx} \right\} - N_\beta \; .
\ea
Note that the sums on the left are over all integer $q$, including $q=0$, but
each has been reorganized into a sum over $q>0$ plus a $q=0$ term,
proportional to the particle number. 

The normal ordering in these expressions is defined so as to the include 
the effect of
the spectral flow, i.e., it is with respect to the vacuum $|0\rangle$ such that
\ba
\alpha_k |0\rangle & = & \beta_k |0\rangle = 0 \; , \;\;\;\;\;\;\; k - g \A_1 > 0\; , 
\\
\alpha_k^\dagger |0\rangle & = & \beta_k^\dagger |0\rangle = 0 \; , 
\;\;\;\;\;\;\; k - g \A_1 \leq 0\; ,
\ea
where $\A_1$ is the constant (zero-momentum) component of the gauge field, and 
$g$ is the gauge coupling. The resulting expressions for the densities are
invariant under the ``large'' gauge transformations.
Once again, we stress that the vacuum thus defined does not have to be 
the true vacuum of the system.

Using the above expressions for the component densities,
we can express the total charge density as
\be
\rho = :\psi_R^\dagger \psi_R: + :\psi_L^\dagger \psi_L: 
= \frac{1}{\sqrt{\pi}} \partial_x \phi + \frac{1}{2\pi} (N_\alpha - N_\beta) \; ,
\label{rho}
\ee
where
\be
\phi = \frac{1}{\sqrt{2\pi}} \sum_{q>0} \frac{1}{\sqrt{2q}} \left\{
(A_q + B_q^\dagger) e^{iqx} + (A_q^\dagger + B_q) e^{-iqx} \right\}
\label{phi}
\ee
is a scalar field. It has not yet acquired a zero ($q=0$) mode but for now, as
only the derivative of $\phi$ occurs in Eq. (\ref{rho}), 
that does not matter. The zero-charge sector is defined by the condition
\be
N_\alpha - N_\beta \approx 0 \; .
\label{zero-charge}
\ee
We use Dirac's $\approx$ notation to indicate that this is a condition imposed on
states, rather than operators. In this sector, Eq. (\ref{cur}) becomes
the $\mu=0$ component of Eq. (\ref{cur}) (for each $f=1,2$ separately).

Similarly, the electric current density can be expressed as
\be
j = :\psi_R^\dagger \psi_R: - :\psi_L^\dagger \psi_L: 
= -\frac{1}{\sqrt{\pi}} p_\phi + \frac{1}{2\pi} (N_\alpha + N_\beta) \; ,
\label{j1}
\ee
where
\[
p_\phi = - \frac{i}{2\sqrt{\pi}} \sum_{q>0} \sqrt{q}
\left\{ (A_q - B_q^\dagger) e^{iqx} + (B_q - A^\dagger_q) e^{-iqx} \right\} \; .
\]
Comparing this to Eq. (\ref{phi}), we see that the Fourier modes of 
$p_\phi$ are canonical momenta conjugate to the (non-zero) modes of $\phi$.
For the zero mode, however, we need to be more careful: our normal ordering of
fermions depends on the gauge field, and so $N_\alpha + N_\beta$ is not the
zero-mode canonical momentum ($p_0$) of the scalar. The correct expression is
\[
N_\alpha + N_\beta = - \frac{1}{\sqrt{\pi}} \left(
p_0 + \frac{g}{\sqrt{\pi}} \Phi \right) \; ,
\]
where $\Phi = \oint \A_1 dx$ is the flux through the circle.
Eq. (\ref{j1}) then becomes the $\mu=1$ component of Eq. (\ref{cur}).

In this notation, the bosonization identities \cite{Delft&Schoeller} read  
\begin{eqnarray}
\psi_R(x) & = & \frac{1}{\sqrt{2\pi}} F_\alpha e^{i(N_\alpha - \delta) x }
:e^{i \sum_{q>0} \frac{1}{\sqrt{q}} [ A_q e^{iqx} + A_q^\dagger e^{-iqx}] }: 
\label{psi_R} \; , \\
\psi_L(x) & = & \frac{1}{\sqrt{2\pi}} e^{i(N_\beta - \delta) x } F_\beta^\dagger
:e^{-i \sum_{q>0} \frac{1}{\sqrt{q}} [ B_q e^{-iqx} + B_q^\dagger e^{iqx}] }: 
\label{psi_L} \; ,
\end{eqnarray}
where $F_\alpha$ and $F_\beta$ are the so-called Klein factors 
(or ``ladder operators''), which are unitary, anticommuting
($\{ F_\alpha, F_\beta \} = \{ F_\alpha, F_\beta^\dagger \} = 0$) and satisfy
these relations:
\[
F_\alpha e^{iN_\alpha \xi} = e^{i(N_\alpha +1) \xi} F_\alpha 
\]
and the same with $\alpha \to \beta$; 
$\xi$ is an arbitrary $c$-number. Again, a simplification occurs in the zero-charge
sector: from Eqs. (\ref{psi_R}) and (\ref{psi_L}) we obtain, for example,
\[
\psi_L^\dagger \psi_R \approx \frac{1}{2\pi} F_\beta F_\alpha :e^{2i\sqrt{\pi} \phi}:
\; ,
\]
where $\phi$ is the field (\ref{phi}).
Moreover, since $F_\beta F_\alpha$ is a unitary Bose operator, we can express it as
$F_\beta F_\alpha = - e^{2i\sqrt{\pi} \phi_0}$, where $\phi_0$ is a single degree of
freedom which now becomes the zero-momentum mode of the field $\phi$.

\end{document}